\documentclass[conference]{IEEEtran}
\IEEEoverridecommandlockouts
\usepackage[utf8]{inputenc}
\usepackage[backend=bibtex,style=ieee,natbib=true,mincitenames=1,maxcitenames=2]{biblatex}
\usepackage{amsmath,amssymb,amsfonts}
\usepackage{mathtools}
\usepackage{graphicx}
\usepackage{textcomp}
\usepackage{xcolor}
\usepackage[hidelinks]{hyperref}
\hypersetup{breaklinks=true}
\usepackage[capitalise]{cleveref}
\usepackage[ruled]{algorithm}
\usepackage[endLComment=]{algpseudocodex}
\usepackage{array,multirow,pbox}
\usepackage{booktabs} 
\usepackage{tabularx}
\usepackage[inline]{enumitem}
\usepackage{xargs}
\usepackage{todonotes}
\usepackage{xspace}
\usepackage{kantlipsum}
\usepackage{comment}
\usepackage{subcaption}

\def\BibTeX{{\rm B\kern-.05em{\sc i\kern-.025em b}\kern-.08em
    T\kern-.1667em\lower.7ex\hbox{E}\kern-.125emX}}

\newcommand{\pic}{\texttt{Picasso}}
\newcommand{\AlgIn}{\Statex \textbf{Input:} }
\newcommand{\AlgOut}{\Statex \textbf{Output:} }

\newtheorem{lemma}{Lemma}

\newtheorem{definition}{Definition}

\def\calL{\mathcal{L}}
\def\calP{\mathcal{P}}

\def\calC{\mathcal{C}}

\def\bard{\bar{d}}
\def\bbE{\mathbb{E}}

\newcommand{\MH}{\textcolor{blue}}

\newcommand{\SMF}[2][inline]{\todo[color=cyan!50,#1]{\sf \textbf{Ferdous:} #2}\xspace}

\newcommandx{\smf}[2][1=]{\todo[color=cyan!50,#1]{\sf \textbf{smf:} #2}\xspace}
\newcommandx{\mm}[2][1=]{\todo[color=orange!50,inline,#1]{\sf \textbf{Marco:} #2}\xspace}

\makeatletter
\newcommand{\crefnames}[3]{%
  \@for\next:=#1\do{%
    \expandafter\crefname\expandafter{\next}{#2}{#3}%
  }%
}

\makeatother
\crefnames{section,subsection}{\S}{\S\S}
\crefname{figure}{Fig.}{Fig.}

\newcommand{\ignore}[1]{{}}

\begin{document}

\title{\texttt{Picasso}: Memory-Efficient Graph Coloring Using Palettes With Applications in Quantum Computing
}

\author{
    \IEEEauthorblockN{S M Ferdous\IEEEauthorrefmark{1},
    Reece Neff\IEEEauthorrefmark{2}\IEEEauthorrefmark{1},
    Bo Peng\IEEEauthorrefmark{1},
    Salman Shuvo\IEEEauthorrefmark{1},
    Marco Minutoli\IEEEauthorrefmark{1},
    Sayak Mukherjee\IEEEauthorrefmark{1},\\
    Karol Kowalski\IEEEauthorrefmark{1},
    Michela Becchi\IEEEauthorrefmark{1}\IEEEauthorrefmark{2},
    Mahantesh Halappanavar\IEEEauthorrefmark{1}}
    \IEEEauthorblockA{
    \IEEEauthorrefmark{1}Pacific Northwest National Laboratory, Richland, WA 
    \IEEEauthorrefmark{2}North Carolina State University, Raleigh, NC\\
    \IEEEauthorrefmark{1}\{FirstName.LastName\}@pnnl.gov\,
    \IEEEauthorrefmark{2}\{rwneff, mbecchi\}@ncsu.edu}
}

\begin{comment}
    
\author{\IEEEauthorblockN{1\textsuperscript{st} SM Ferdous}
\IEEEauthorblockA{\textit{dept. name of organization (of Aff.)} \\
\textit{Pacific N}\\
City, Country \\
email address or ORCID}
\and
\IEEEauthorblockN{2\textsuperscript{nd} Given Name Surname}
\IEEEauthorblockA{\textit{dept. name of organization (of Aff.)} \\
\textit{name of organization (of Aff.)}\\
City, Country \\
email address or ORCID}
\and
\IEEEauthorblockN{3\textsuperscript{rd} Given Name Surname}
\IEEEauthorblockA{\textit{dept. name of organization (of Aff.)} \\
\textit{name of organization (of Aff.)}\\
City, Country \\
email address or ORCID}
\and
\IEEEauthorblockN{4\textsuperscript{th} Given Name Surname}
\IEEEauthorblockA{\textit{dept. name of organization (of Aff.)} \\
\textit{name of organization (of Aff.)}\\
City, Country \\
email address or ORCID}
\and
\IEEEauthorblockN{5\textsuperscript{th} Given Name Surname}
\IEEEauthorblockA{\textit{dept. name of organization (of Aff.)} \\
\textit{name of organization (of Aff.)}\\
City, Country \\
email address or ORCID}
\and
\IEEEauthorblockN{6\textsuperscript{th} Given Name Surname}
\IEEEauthorblockA{\textit{dept. name of organization (of Aff.)} \\
\textit{name of organization (of Aff.)}\\
City, Country \\
email address or ORCID}
}
\end{comment}

\maketitle
\thispagestyle{plain} %%Remove before submission
\pagestyle{plain}

%%%
\begin{abstract}
A \emph{coloring} of a graph is an assignment of colors to vertices such that no two neighboring vertices have the same color. The need for memory-efficient coloring algorithms is motivated by their application in computing clique partitions of graphs arising in quantum computations where the objective is to map a large set of Pauli strings into a compact set of unitaries. We present \texttt{Picasso}, a randomized memory-efficient iterative parallel graph coloring algorithm with theoretical sublinear space guarantees under practical assumptions. The parameters of our algorithm provide a trade-off between coloring quality and resource consumption. To assist the user, we also propose a machine learning model to predict the coloring algorithm's parameters considering these trade-offs. We provide a sequential and a parallel implementation of the proposed algorithm.

We perform an experimental evaluation on a 64-core AMD CPU equipped with 512 GB of memory and an Nvidia A100 GPU with 40GB of memory. For a small dataset where existing coloring algorithms can be executed within the 512 GB memory budget, we show up to {\bf 68$\times$} memory savings. On massive datasets we demonstrate that GPU-accelerated \pic{} can process inputs with {\bf 49.5$\times$} more Pauli strings (vertex set in our graph) and {\bf 2,478$\times$} more edges than state-of-the-art parallel approaches.

%To the best of our knowledge, this is the first paper to develop a parallel algorithm and provide empirical analysis with up to $2$ million vertices and over a trillion edges. The current graph algorithms are limited by memory constraints and are not able to solve such large instances. 
%\SMF{Mention Kokkos, ECL-GC comparison results.}
\end{abstract}

\section{Introduction}
\label{sec:Intro}

Given a graph $G(V,E)$, the problem of graph coloring is to assign a color to each vertex such that no two adjacent vertices are assigned the same color, while minimizing the number of colors used.
Graph coloring is one of the central problems in combinatorial optimization with applications in various scientific domains~\cite{Lewis2015}.  Many algorithmic techniques have been developed to solve coloring in sequential, parallel, and distributed settings for many variants of the coloring problem~\cite{Gebremedhin2005}. All of these algorithms are memory-demanding since they require loading the entire graph and several auxiliary data structrues into the memory.  
%Most of them require loading the entire graph and several auxiliary data structures into memory and, therefore, they are avid of memory. 
For massive graphs, especially on limited-memory accelerators (GPUs), these algorithms easily exhaust the available memory, and the problem is exacerbated for dense graphs due to quadratic scaling of edges. Our work considers graphs that are $\approx50\%$ dense ($|E| \approx |V|^2/2$) for which the current graph coloring approaches \citep{Gebremedhin2013, deveci2016parallel, Alabandi2022, Rajamanickam2021} quickly run out of memory when processing large instances (\cref{sec:ExptResults}).

Our quest for memory-optimized coloring techniques stems from an application of quantum algorithms to computational chemistry. Quantum computers have the potential to offer new insights into chemical phenomena that are not feasible with classical computers. To utilize quantum approaches effectively, it is indispensable to encode chemical Hamiltonians and chemical-inspired wave function ans\"{a}tze (i.e., the assumptions of the wave function form) in a quantum-compatible representation, while adhering to the proposed theoretical framework. However, the direct encoding of chemical Hamiltonians and typical chemical-inspired ans\"{a}tze, which usually grows as high-degree polynomials, is unscalable. This directly affects the efficiency and applicability of the corresponding quantum algorithms. For example, transforming the chemical Hamiltonian from the second quantization to spin operators yields \emph{Pauli strings} (tensor products of 2 $\times$ 2 Pauli and identity matrices) that scale as $\mathcal{O}(N^4)$, where $N$ is the number of basis functions spanning the Hamiltonian~\cite{Kandala2017hardware,mcclean2016theory,Izmaylov2020}. 

One near-term solution to improve the scaling of quantum algorithms is the \emph{unitary partitioning} and its variants, which aim to find compact representations of a linear combination of Pauli strings (\cref{sec:problem}). These strategies, predominantly based on the graph analysis of Pauli strings, typically yield a reduction ranging from 1/10 to 1/6 in the number of alternative unitaries for small test cases. The unitary partitioning problem reduces to a graph coloring problem, where the vertices of the graph represent the Pauli strings, and the edges represent whether these  strings obey an anticommute relation (\cref{sec:problem}). These graphs are large and dense. Therefore, the state-of-the-art approaches for the unitary partitioning problem can only solve small molecules with a few thousand Pauli strings, whereas the desired scale is on the order of $\mathcal{O}(10^6\sim10^{12})$ of Pauli strings.

Building on recent developments in sublinear algorithms for graph coloring~\cite{AssadiCK19}, we introduce a novel parameterized coloring algorithm, \pic{}, where the parameters of the algorithm provide a trade-off between the quality of the solution (i.e., the number of colors used) and the resource consumption of the algorithm. We theoretically prove that \pic{} has a sublinear space requirement (\cref{sec:Algorithm}), making it attractive to implement on GPUs. We empirically show that \pic{} can solve {\em trillion edge graph problems} arising from molecules with more than 2 million Pauli strings in under fifteen minutes. Many of the results we report are the first-of-its-kind for the corresponding molecule and basis set. Since our parametric algorithm considers the target number of colors, we show that high-quality coloring 
%compared to the state-of-the-art 
can be achieved by an aggressive choice of parameters (\cref{subsec:quality-mem}). 
%To assist the users in determining the parameters, 
We design a machine learning prediction model to determine the parameters configuration to be used to achieve a given trade-off between coloring quality and resource requirements (\S\ref{sec:Prediction}). 

Although \pic{} is designed to solve a specific problem in quantum computing, it can be used in a generalized graph setting where memory efficiency is needed. The main contributions of this work are:
\begin{itemize}%[leftmargin=*, noitemsep, topsep=2pt]
%\begin{itemize}[ noitemsep, topsep=2pt]
    \item We introduce a first-of-its-kind graph coloring algorithm \pic{} to address the unitary partitioning problem in quantum computing with demonstrations for dense inputs with up to two million vertices and over a {\em trillion edges}.
    \item We prove that, under a practical assumption \pic{} has {\em sublinear memory} requirement with high probability.
    % \item We employ a machine learning approach to predict the parameters of our algorithm that can optimize for a minimal number of colors, runtime, and memory footprint for specific application needs and resource constraints.
    \item We propose a machine learning approach to predict the configuration of the algorithm's parameters that allows achieving a given trade-off between coloring quality, runtime and memory requirements
    %\item For the small molecule systems that can be solved by existing algorithms with a 512 GB memory budget, we show up to {\bf 68$\times$} memory reduction.
    %\item We port the most compute-intensive portion of \pic{} to GPU and demonstrate up to {\bf $173\times$} speed-up relative to single-threaded versions for small problems.
    \item We demonstrate the practical efficiency of our approach. The GPU-accelerated \pic{} enables to process inputs with {\bf $49.5\times$} more Pauli strings and {\bf $2,478\times$} more edges than existing state-of-the-art parallel approaches.
    %\item Demonstrate up to $100 \times$ reduction in memory footprint.
\end{itemize}

\section{Problem Formulation}
\label{sec:problem}
Our work is driven by the challenge of identifying compact unitary representations of chemical Hamiltonians and strongly correlated wave functions to enable accurate and efficient quantum simulations. Typically, this challenge can be abstracted as determining how to solve a clique partitioning problem efficiently. In this section, we delve into the graph formulation of the problem.

%%%%%%%%%%%%%%%%%%%%%%%%%%%%%%%%%%%%
\begin{figure*}[!htb]
\centering
\includegraphics[width=0.90\textwidth]{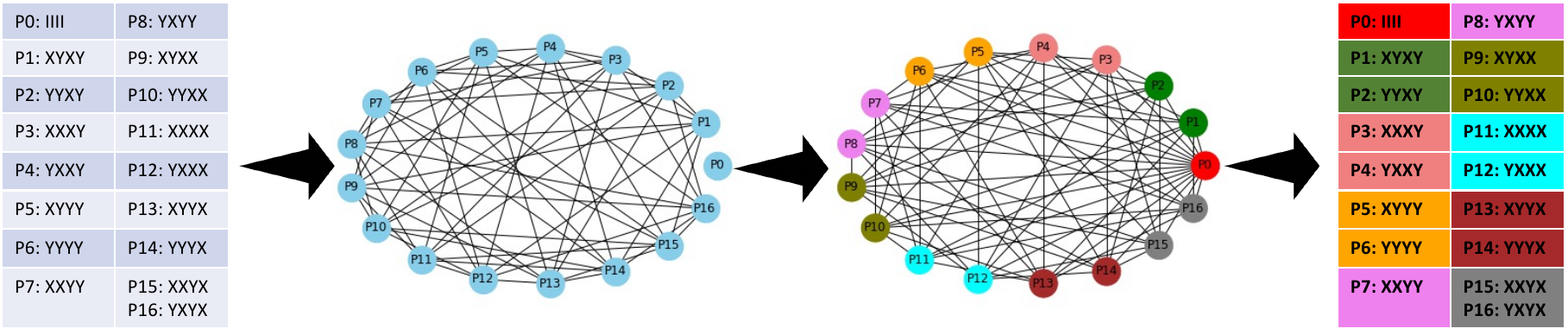} 
\caption{An overview of the mapping problem solved as clique partition using graph coloring of the conflict graph for H2 molecule with sto-3g basis function.}
\label{fig:Overview} 
\end{figure*}
%%%%%%%%%%%%%%%%%%%%%%%%%%%%%%%%%%%%

\subsection{Quantum computing problem}
In quantum simulations aimed at elucidating physical and chemical phenomena, both the wave function and the system Hamiltonian play crucial roles. The wave function encodes the probabilistic state of a quantum system, facilitating the computation of various physical properties, whereas the system Hamiltonian governs the temporal evolution of the system's states according to quantum mechanical principles and represents the total energy of the system. In this context, the wave function is responsible for state preparation, and the system Hamiltonian controls the evolution of the state.

In these simulations, the processes of state preparation and evolution \emph{must} be conducted through \emph{unitary operations}. These are operations that maintain the norm (magnitude) of the state vector in Hilbert space, ensuring that the total probability remains one. 
To accomplish this, both the system Hamiltonian and the wave function generator must be reformulated as either a unitary operator or a linear combination of unitary operators. This reformulation is crucial because unitary operations are reversible and preserve the quantum information within the system, making them indispensable for the coherent manipulation of quantum states in simulations.
%\sayak{can we please give some basic definitions of the quantum quantities} 
In small-scale demonstrations, one could employ, for instance, the Jordan-Wigner, Bravyi-Kitaev, or parity techniques~\cite{JW1928,bravyi2002Fermionic}, which rewrite the Hamiltonian and the wave function generator as a combination of Pauli strings, each being a unitary. A Pauli string represents a tensor product of a series of $2 \times 2$ Pauli matrices, $\sigma_x,\sigma_y,\sigma_z$, and the $2\times2$ identity matrix, $I$. However, these techniques encounter the curse of dimensionality in large-scale applications. For instance, in molecular applications, reformulating the molecular Hamiltonian represented in the $N$-spin-orbital basis set would require $\mathcal{O}(N^4)$ Pauli strings. The number of the Pauli strings needed to reformulate the highly correlated wave function ans\"{a}tze, such as the non-unitary coupled-cluster ans\"{a}tz that include single and double excitations, would be even higher, i.e. $\mathcal{O}(N^{7\sim8})$. The curse of dimensionality makes the subsequent quantum simulation, especially quantum measurements, extremely challenging. Compact unitary representation for both the Hamiltonians and the wave function generators addresses the curse of dimensionality and the associated quantum measurement overhead. 

%algorithm such as Variational quantum eigensolver (VQE) [refernces] and the Quantum Method of moment coupled-cluster equation(MMCCQ)\cite{peng2022mapping} techniques, we are provided with a non-unitary operator which must have to be represented by a set of unitaries since quantum computers can only accept unitary forms. An arbitrary non-unitary operator can be transformed into a linear combination of Pauli terms (a multiplication of a set of $2\times2$ unitary pauli matrices). The number of pauli terms in the linear presentation grows large($~N^4$ in VQE methods), where $N$ is the size of the system. To optimize the quantum measurement, one idea is to group the pauli terms with special properties so that they can be measured simultaneously[refernce]. 
Mathematically, given a set of $n$ Pauli strings, $P = \{P_1,P_2, \ldots, P_n\}$ each of which is of length $N$, along with their coefficients $\{p_1,p_2,\ldots, p_n\}$, we aim to compute a smaller set of unitaries $\{U_1, U_2, \ldots,U_c\}$ with the corresponding coefficients $\{u_1,u_2,\dots, u_c\}$, such that:
\begin{align}
    \sum_{i=1}^c u_iU_i = \sum_{j=1}^n p_jP_j, ~~ c<n. \label{eq:compact}
\end{align}
%\MH{Bo: ``$c$-unitary'' is incomplete -- is it ``$c$ unitaries''?}

\subsection{Connection to Clique partitioning and Graph coloring}
A straightforward way for satisfying Eq. (\ref{eq:compact}) is enabled through the following condition for 
%general 
$m$ Pauli strings~\cite{peng2022mapping}:
\begin{align}
    \label{eq:group-cond}
    \sum_{i,j=1,i \neq j}^m p_i^*p_jP_iP_j = 0, \ \ m \ge 2,
\end{align}
for which a necessary condition is enabled by the \emph{anticommute property} between any two Pauli strings in the given set, i.e.,
\begin{align}
    \{p_i P_i, p_j P_j\} = p_i^* p_j P_i P_j + p_j^* p_i P_j P_i = 0, ~~ i\neq j. \label{eq:edge-cond}
\end{align}
The above relationships can be translated to a \emph{clique partitioning} and \emph{coloring} problem in a graph as follows. Given a set of Pauli strings, $P$, we generate a graph $G(P,E)$, where $E$ is the set of all possible (unique) pairs of Pauli strings in $P$ that anticommute, i.e., satisfy the Eq. (\ref{eq:edge-cond}). A \emph{clique} (or a complete subgraph) of a graph is a set of vertices such that every (unique) pair of vertices in that set are connected with an edge in $G$. A clique in $G$ thus corresponds to satisfying Eq. (\ref{eq:group-cond}). Our goal is to generate a set of cliques, as small as possible, in $G$, which forms a partition of $P$. We formally define the problem as follows.

%%%%%%%%%%%%%%%%%%%%%%%%%%
\begin{definition}[Clique partitioning]
Given a set of Pauli strings, $P = \{P_1,P_2,, \ldots, P_n\}$, and a graph $G(P, E)$ generated from $P$, the clique partitioning problem is to compute a collection of cliques, $U= \{U_1,\ldots, U_c\}$, where $\bigcup_i U_i = P$, and $U_i \cap U_j = \emptyset$, for $i \neq j$ that minimizes the number of cliques, i.e., $c$. 
\end{definition}
%%%%%%%%%%%%%%%%%%%%%%%%%%%%%%%%%

The clique partitioning problem is NP-Complete~\cite{GareyJ79}, which makes any optimal polynomial time algorithm unlikely. 
%The clique partition 
This problem is equivalent to the well-known \emph{graph coloring} problem, which requires finding the smallest set of colors such that each vertex of the graph is assigned exactly one color from the set, and no two endpoints of an edge have the same color. The clique partitioning of a graph $G$ is equivalent to the coloring of the complement graph $G^\prime$~\cite{bhasker1991clique}. 
%To see the connection between clique partitioning and coloring, 
Let $G^\prime(V,E^\prime)$ be the complement of a graph $G(V,E)$, where the vertex set remains the same but the edge set, $E^\prime = \{\{V \times V\} \setminus E\} \setminus \{\{v,v\}: v \in V\} $. 
Since the vertices of $G^\prime$ with the same color must represent a clique in $G$, it is easy to verify that any proper coloring of $G^\prime$ provides a feasible solution to the clique partitioning of $G$. We further note that when $|E| \geq |E^\prime|$, computing clique partitioning via coloring of the complement graph is more efficient. \looseness = -1

In this paper, we solve the unitary partitioning problem by formulating it as a clique partitioning problem, which in turn is solved using graph coloring on the complement graph. An illustration of the process is shown in Fig.\ref{fig:Overview} using 
%, we show a workflow of 
the clique partitioning  to generate a compact representation of the H2 molecule with sto-3g basis function, with $N=4$, as an example. We begin with a set of Pauli strings as vertices and compute the edges of the graph using Eq.\ref{eq:edge-cond}. We then construct the complement graph and color it. Finally, we output the partition according to the color classes. In the example, 17 Pauli strings are shrunk to a set of 9 unitaries. 
\section{Related Work}

\noindent \textbf{Mapping Pauli Strings:} To facilitate efficient term-by-term measurement schemes in NISQ devices, efforts to minimize the number of terms representing the Hamiltonian and wave function generators have been essential. Previous developments in hybrid quantum-classical algorithms have primarily focused on classically grouping Pauli strings to reduce quantum measurements. Drawing inspiration from the concept of Mutually Unbiased Bases (MUB) in quantum information theory \cite{Schwinger1960unitary,klappenecker2004mub}—which is associated with maximizing the information gained from a single measurement—early endeavors have leveraged the commutativity of Pauli strings. This includes qubit-wise commutativity~\cite{Altepeter2004}, general commutativity~\cite{gokhale2020,Izmaylov2020}, and unitary partitioning~\cite{izmaylov2019unitary}, to define the edges when translating the grouping problem to a clique partitioning and coloring problem in a graph composed of these Pauli strings.

For general molecular cases explored in quantum simulations, these strategies have the potential to group $(4^{N_q}-1)$ $N_q$-qubit Pauli strings (excluding the identity string) into no more than $3^{N_q}$ groups, with further reductions to $\mathcal{O}(N_q^{2\sim 3})$ possible, albeit at the cost of introducing additional one/multi-qubit unitary transformations before measurement. However, the polynomial scaling inherent in these grouping schemes renders script-based large-scale applications unscalable. This necessitates the development of high-performance computing libraries specialized in graph analysis for this purpose.

\noindent \textbf{Graph Coloring:}
%As a prototypical graph problem, 
Graph coloring has been studied extensively in literature. The application of graph coloring in automatic (or algorithmic) differentiation has led to the study of different types of graph coloring problems 
%, which is well described in the excellent survey by \citet{Gebremedhin2005}
 \cite{Gebremedhin2005}, and a software library of serial implementations called ColPack~\cite{Gebremedhin2013}. Sequential coloring algorithms are based on \emph{greedy} methods, which, given an ordering of the vertices, employ the smallest feasible color for each vertex. In the worst case and for general graphs, all these ordering-based methods require $\Delta + 1$ colors, where $\Delta$ is the maximum degree of the graph. The ordering methods include Largest Degree First (LF), Smallest Degree Last (SL), Dynamic Largest Degree First (DLF), and Incidence Degree (ID). 
%Since coloring is a well-known NP-hard problem, heuristics have received more attention than exact approaches. 
%In their survey~\cite{Lima2018}, de Lima  and Carmo group exact approaches into Dynamic Programming, Branch-and-Bound~\cite{Brelaz1979}, and Integer Linear Programming. 
While greedy coloring techniques provide reasonable quality in sequential settings, these algorithms have little to no concurrency in practice. For parallel settings, there are two primary algorithmic techniques: 
\begin{enumerate*}[label=\textit{\roman*)}]
    \item heuristics building on the idea of finding maximal independent sets, introduced in the pioneering work of \citet{Luby1986} and extended by \citet{Jones1993} (JP),
    \item heuristics leveraging {\em speculation}, where parallel threads speculatively color vertices using the least available color.
\end{enumerate*}
Conflicts resulting from concurrent execution are then corrected in an iterative manner~\cite{Catalyurek2012}. 
%\SMF{Need to expand the next paragraph to discuss several gpu coloring algorithms, especially the two we are comparing}
Similar approaches have been extended to manycore or GPU implementations~\cite{Nguyen2018, Bogle2020, Alabandi2022}, and distributed-memory algorithms and implementations~\cite{Catalyurek2011, Bozdag2010}. In
\cref{sec:ExptResults} we compare \pic{} for quality (against ~\cite{Gebremedhin2013}) and performance (against ~\cite{Bogle2020,Alabandi2022}). \looseness = -1

%\MH{Ferdous, please complete this portion:}
The existing parallel graph coloring algorithms, especially the single-node GPU solutions, fail to solve large graph problems due to memory limitations. The graphs need to be loaded into the GPU memory along with auxiliary data structures such as an array of ``forbidden colors'', which can easily exhaust the available memory. The graphs generated by our target applications are large and dense. Thus, there exists a need for coloring algorithms that are memory efficient. %Also many high performance coloring algorithms are only designed for relatively sparse graphs. For example, both the multi-threaded and gpu coloring implementation of ECL-GC~\cite{AlabandiB22}, could not find a valid coloring for the graphs in our dataset.  

%Most of the works discussed so far are designed for graph instances that are relatively sparse in nature. For instance, memory requirements to store forbidden colors in a parallel algorithm can go beyond available memory of the system. Consequently, our approach aims to minimize the memory footprint by trading off with quality (\# of colors used) and runtime (relatively slower). We build on the theoretical insights from X and Y in using a palette of colors of a certain size. To the best of our knowledge, this is the first-of-its-kind work that demonstrsates coloring of dense graphs with millions of vertices on GPUs.

\begin{comment}
\begin{itemize}
    \item Clique-partition literature
    \item Sequential graph coloring: ordering based
    \item Parallel graph coloring: JP, speculative
    \item GPU graph coloring
    \item why all those fails in our application?
\end{itemize}
\end{comment}

\noindent \textbf{Graph Coloring in sublinear space}: The recent seminal work of Assadi, Chen and Khanna~\cite{AssadiCK19} (ACK, henceforth), stduied the $(\Delta+1)$-coloring problem with sublinear space and time constraints. They developed the \emph{Pallete Sparsification Theorem}, which reduces the $\Delta+1$-coloring of a graph $G(V,E)$ to a list-coloring problem in a subgraph of $G$ with only $\mathcal{O}(|V| \log^2{|V|})$ edges. Applications of this theorem is shown by designing algorithms in dynamic semi-streaming in a single pass, sublinear query and MPC models. In terms of semi-streaming algoirthm, the only known $(\Delta+1)$-coloring before ACK's algorithm~\cite{AssadiCK19} was the $\mathcal{O}(\log{|V|})$ pass distributed algorithm of Luby~\cite{Luby1986,luby1993removing}, simulated in streaming setup. In semi-streaming model, ACK uses palette size as $\Delta+1$, list size of colors as $\mathcal{O}(\log |V|)$ for each vertex, and a special post-processing step. In this paper, we put ACK's algorithm in practice by non-trivial modifications listed as follows.

\begin{enumerate}[label=\textit{\roman*)}]
    \item The palette size in ACK's algorithm is $\Delta+1$, which limits the practicality of the algorithm on problems operating on large dense graphs, like the one we considered. Our graphs (original and complement) are dense ($\Delta > |V|/2$). One would require a fraction of $|V|$ to color all the vertices, as shown in \cref{tab:quality-compare} in \cref{subsec:quality-mem} for our dataset ($<=16$\% of $|V|$). Our algorithm allows users to specify a variable palette size. Modifying the analysis of~\cite{AssadiCK19}, we proved that if the ratio $\Delta/\calP$ is bounded (by $\log{|V|}$), a sublinear space is guaranteed. This assumption holds for our graphs since this ratio is a constant for our use cases.
    \item The conflict graph coloring algorithm of~\cite{AssadiCK19} decomposes the graph and then applies a greedy coloring, an almost clique coloring, and a maximum matching-based coloring on the decomposition. We provide an efficient implementation of list-greedy-based coloring that dynamically colors the vertices based on their color list size.
    \item  ACK's streaming algorithm is single-pass. For valid coloring, the single iteration algorithm would require a large palette size, degrading the solution quality. We address this issue with an iterative approach, where in each iteration, we attempt to color the uncolored vertices from the previous iteration. Our proof for sublinear space holds for each iteration.
    
\end{enumerate}

\section{Our Algorithm}
\label{sec:Algorithm}
%In this section, we present \pic{}, a meta-algorithm for coloring, listed in Algorithm~\ref{alg:pal-col-meta}. We then discuss implementations of the algorithm. %\sayak{In the previous section complement graph is denoted by $G'$}

%The algorithm first assigns a random list of colors to each vertex and computes the \emph{conflict subgraph}, which consists of the edges that share a common color in their corresponding lists. 
%\pic{} is inspired by the seminal ACK's streaming algorithm~\cite{AssadiCK19}. By choosing the size of the palette as $\Delta+1$, the size of the color list as $\mathcal{O}(\log |V|)$, and adopting a special post-processing step, ACK's algorithm can color a graph with maximum degree $\Delta$ with $\Delta+1$ colors in a single pass in sublinear space and time. We extend and modify this algorithm in several ways:

%We note that the $\Delta+1$ bound is pessimistic for many real-world graphs. Motivated by this observation, \pic{} extends~\cite{AssadiCK19} in several ways: 

 % We experimentally found that this iterative approach achieves better coloring quality than the single pass algorithm. We describe the algorithm in the following narrative. 

%%%%%%%%%%%%%%%%%%%%%%%%%%%%%%%%%%%%%%%%
\begin{table}[!htbp]
    \centering
    \setlength\tabcolsep{2pt} % default value: 6pt
    \caption{Notation used in the paper.} 
\label{tab:notation}
\begin{tabular}{l|l}
\toprule
Symbol & Description \\
\midrule
 $P$ & Set of Pauli strings\\ \hline
  $N$ & Length of each Pauli string\\ \hline
  $\ell$ & Iteration counter\\ \hline
 $G_\ell=\{ V, E \}$ & Complement graph from $V \subseteq P$ at it. $\ell$ \\ \hline
 $n$& number of vertices, $n \coloneqq |V|$ \\ \hline
 $G_c=\{ V_c, E_c \}$ & Conflict graph, $V_c \subseteq V$ \\ \hline
 $\delta(v)$ & Degree of vertex $v$, with neighbor set: $adj(v)$\\ \hline
 $\mathcal{P}$ & Palette size, Palette=$\{(\ell-1)\calP,1,\ldots, \ell\calP-1\}$\\ \hline
 $\mathcal{C}$ & Final number of colors, $\mathcal{C} \leq \mathcal{P}$ \\ \hline
 $\alpha$ & Multiplicative factor for List size \\ \hline
 $\mathcal{L}$ & List Size, set to $\alpha \log |V|$ \\ \hline
 $\beta$ & Weighting factor for bi-objective optimization \\ 
 
\bottomrule
\end{tabular}
    \label{tab:results}
\end{table} 
%%%%%%%%%%%%%%%%%%%%%%%%%%%%%%%%%%%%%%%%

\begin{algorithm}
    \caption{\texttt{Picasso}: Palette-based graph coloring}
    \label{alg:pal-col-meta}
    \begin{algorithmic}[1]
        \AlgIn{A graph $G=(V,E)$}
        \AlgOut{A $color$ array, with a valid coloring of $G$ } 
        \State Initialize the $color$ array
        \State $\ell = 1$ \Comment{The iteration number}
        \State $G_\ell\gets$ $G$ 
        \While { $V$ is not empty }
            \State $(\calP_\ell, \calL_\ell)\gets$ Initialize palette and list size for $G_\ell$ \label{line:est}
            \State $colList \gets$  assign\_rand\_list\_colors($\calP_\ell,\calL_\ell,G_\ell$) \label{line:assgn}
            \State $G_c \gets$ construct\_conflict\_graph(colList, $G_\ell$) \label{line:buildgraph}
            \State color\_unconflicted\_vertices($V$ $\setminus$ $V_c$, $colList$, $colors$) \label{line:color-unconflicted}
            \State $V_u \gets$ color\_conflict\_graph($G_c$, $colList$, $colors$)\label{line:conf-resolve}
            \State $\ell = \ell + 1$
            \State $G_\ell\gets$ subgraph induced by $V_u$ in $G$\label{line:subgraph}
            \State $V\gets V_u$
        \EndWhile
    \end{algorithmic}   
    \label{alg:palette-coloring}
\end{algorithm}

\pic{}, listed in Algorithm~\ref{alg:pal-col-meta}, attempts to color the graph by initially assigning to each vertex a list of candidate colors chosen uniformly at random from a palette of $\calP$ colors $\{0,1,\ldots, \calP-1\}$, and progressively assigning to each vertex a color from its color list without violating the graph coloring constraint. It takes a graph $G$ as input and computes a valid coloring of $G$ in the $color$ array. The algorithm starts with the original graph $G$, and iteratively computes coloring for a subgraph. In each iteration ($\ell$), it estimates the palette size ($\calP_\ell$) and list size ($\calL_\ell)$ for the current subgraph $G_\ell$
(Line~\ref{line:est}). Next, it assigns to each vertex a list of candidate colors chosen uniformly at random from the palette (Line~\ref{line:assgn}). For each vertex, we record the list of colors in the $colList$ data structure, where $colList(u)$ refers to the list of colors assigned to vertex $u$. We build the conflict subgraph, which consists of the edges that share a common color in their corresponding lists (Line~\ref{line:buildgraph}) using the $colList$ array. We then color the unconflicted vertices using an arbitrary color from their color list (Line~\ref{line:color-unconflicted}), and attempt to color the conflict graph (Line~\ref{line:conf-resolve}). The vertices uncolored in the current iteration are denoted as $V_u$. We then compute the subgraph induced by $V_u$ and continue if $V_u$ is non-empty. We note that the algorithm in every iteration starts with a new palette of colors and attempts to color the subgraph with these new colors. The colors of an iteration are not reused in the subsequent iterations. We can ensure that by defining the palette set as $\{(\ell-1)\calP,\ldots \ell\calP\}$ at iteration $\ell$.  % We then use the We call this a meta-algorithm since we need to clarify how the individual procedures are implemented that we describe next. %While assigning random list colors and building the conflict graphs are straightforward, we have several choices for the conflict resolution steps. Here, we discuss two approaches. 

We now describe the construction and coloring of the conflict graph at iteration $\ell$. For ease of presentation, we omit the subscript $\ell$.
%\subsection{Building the conflict graph}
\subsection{Conflict Graph Construction}
\label{sec:pauliencoding}
%The conflict graph consists of the conflicting edges. 
An edge $(u,v)$ in $G$ is conflicted if its two endpoints share a color, i.e., $colList(u) \cap colList(v) \neq \emptyset$. If the color values in $colList$ are sorted, we need $O(\calL)$ time to check for a conflict edge. In our application, we are not provided with the graph. Instead, we are given a set of Pauli strings $P$ that defines the vertex set of $G$. We use a bit encoding scheme to dynamically derive the edges from Pauli strings in a memory-efficient way.
%\subsubsection{Pauli String Encoding}
%\SMF{@Bo: Please see the following paragraph and add a brief discussion on why the following technique is equivalent of solving Eq.~\ref{eq:edge-cond}}
Given two Pauli strings $P_i$ and $P_j$, whether an edge exists between them (in our case, a non-edge) can be checked using Eq.~\ref{eq:edge-cond}. Here, a Pauli string consists of $N$ characters, where each character corresponds to a $2 \times 2$ Pauli matrix:
\begin{align}
    \sigma_x = \left( \begin{array}{cc} 0 & 1 \\ 1 & 0 \end{array} \right),
    \sigma_y = \left( \begin{array}{cc} 0 & -i \\ i & 0 \end{array} \right),
    \sigma_z = \left( \begin{array}{cc} 1 & 0 \\ 0 & -1 \end{array} \right).
\end{align}
When directly implementing the anti-commute condition given by Eq.~\ref{eq:edge-cond}, we must carry out $N-1$ tensor products for each string, followed by two matrix multiplications for matrices $P_i$ and $P_j$. However, the inherent properties of Pauli matrices allow us to check this condition efficiently. Specifically, any pair of Pauli matrices will either commute or anti-commute. Importantly, two distinct Pauli matrices will anti-commute
\begin{align}
    \{\sigma_i,\sigma_j\} = 0 ~~\text{if}~~ \sigma_i \neq \sigma_j \neq I,
\end{align}
which is simplified to an element-wise character comparison.

The anti-commutation property can be extended to Pauli strings. This is done by counting the element-wise character comparisons corresponding to the anti-commute relation. It is worth noting that the anti-commutation relationship between two Pauli strings is determined by phase $\pm i$. Only an odd number of element-wise anti-commuting character comparisons will result in a non-vanishing phase.
%
%In practice, we can make $N$ element-wise comparison between $P_i$ and $P_j$, and maintain a counter initialized to zero. If in any position there is an identity matrix (`$I$') or the two character matches, we do not increment the counter. Otherwise we increment it. Finally after the element-wise comparison, if the counter is odd, we return this as edge, otherwise a non-edge. 

Further reduction in the number of comparisons can be achieved using bit representations.
%\textbf{Bo and/or Ferdous: Please discuss the Pauli string comparison step.} \textcolor{blue}{Bo: this looks cool to me. We can refine it later once the initial population of each section is done.}
%During Pauli string comparison, each Pauli matrix from one string is compared with the corresponding matrix in the other string, resulting in $N$ operations. To reduce the number of comparison operations, 
Specifically, we make an observation: as there are only four possible matrices ($\sigma_x, \sigma_y, \sigma_z, I$), we can encode an 8-bit character datatype to a smaller representation using bits. If we encode each Pauli matrix into a 2-bit value, we would still need to perform $N$ comparisons to count the number of mismatches.  To determine a complement edge, we only check whether the number of mismatches is even or odd, which is determined by observing the least significant bit of the count value. With this, we need an encoding that causes a bit flip only when there is a mismatch between $\sigma_x, \sigma_y, $ or $\sigma_z$.

To accomplish this, we implement an encoding scheme similar to an inverse one-hot encoding, where matrices $\sigma_x, \sigma_y,$ and $\sigma_z$ are assigned 110, 101, and 011, and $I$ is assigned 000, respectively. We perform a bitwise \texttt{AND} operation between the encoded Pauli strings and perform a \texttt{popcount} operation to count the number of `1' bits. Due to this encoding scheme, the only time the least significant bit is flipped in the \texttt{popcount} is during a mismatch, allowing us to track whether there is an odd or an even number of mismatches. Speedups from the encoded implementation on CPU range from $1.4$ to $2.0\times$, including the encoding overheads.
%Assuming we encode 10 Pauli matrices to a 32-bit integer, this method results in 30x fewer comparisons with the added overhead of popcount, where the implementation is architecture-specific. In addition, 
%The number of memory accesses is reduced by $2.67\times$ when the 8-bit characters are encoded into 3-bit values.
%\SMF{What is meant by number of memory accessed? does it reduce memory to the overall algorithm?}

%\subsection{Properly list-color the conflict graph}
\subsection{Coloring the Conflict Graph}
Once we compute the conflict graph, $G_c$ we are required to color $G_c$ using the list of colors assigned to each vertex. Here, we discuss two possible approaches to achieve that.

\noindent \textbf{Static order schemes:}
%\paragraph*{Static order schemes}
Given an ordered set of vertices, we iterate through the set and attempt to color each vertex with the first available color in its list that does not conflict with the colors already assigned to its neighbors. We may use popular~\cite{Gebremedhin2005} vertex order such as the Natural, Largest Degree First, Smallest Degree Last, or Random ordering.

\noindent\textbf{Dynamic vertex order scheme:}
A second approach is to color the conflict graph using the lists in a dynamic order~\cite{AchlioptasM97}. Here, we describe a dynamic greedy algorithm from~\cite{AchlioptasM97} for the list coloring and discuss an efficient implementation of it. The algorithm attempts to color the vertices that are most constrained, i.e., it colors a vertex according to the size of the list. %The vertex with the list with the smallest size is colored first; removing this color from the neighboring vertices changes the ordering dynamically. 
When a vertex is assigned a color from its list, the assigned color is removed from the list of all the neighbor of the vertex, rendering a dynamic order on the vertices. 
A na\"ive implementation of the dynamic list coloring algorithm would require $\mathcal{O}(|V_c|^2 +|E_c| \calL)$ time. 
We have at most $|V_c|$ iterations, and in each iteration, we need to find the vertex with the smallest current list size in $\mathcal{O}(|V_c|)$ time, then color the vertex and mark this color removed from all its neighbors in $\mathcal{O}(\delta(v)\calL)$ time, where $\delta(v)$ is the degree of vertex $v$. Summing over $|V|$ gives us the total time. 
We can reduce it to $\mathcal{O}((|V_c| +|E_c| \calL) \log{|V_c|})$ using a minimum heap data structure with logarithmic \texttt{update\_key} operation. Next, we present an efficient implementation in Algorithm~\ref{alg:list-color-dyn-grd} that eliminates the $\log{|V_c|}$ factor by using the bucketing technique. 

\begin{algorithm}
    \caption{Greedy list-coloring of conflict graph, $G_c$}
    \label{alg:list-color-dyn-grd}
    \begin{algorithmic}[1]
        \AlgIn{Conflict graph $G_{c}$, color list $colList(v)$, $\forall v \in V_c$.}
        \AlgOut{A coloring of $G_{c}$ using the lists, and the uncolored vertex set, $V_u$.}
        \State $B \gets$ An array of bucket lists %\MH{Is this an empty bucket list?}
        \LComment{Creating the initial buckets}
        \For{ $v \in V$}
            \State Insert $v$ to $B[colList(v).size()]$
        \EndFor
        \State Mark all vertices of $G_c$ as unprocessed
        \State $V_u \gets \emptyset$
        \While{ $\exists$ an unprocessed vertex}
            \State Pick a vertex, $v$ from the lowest bucket 
            \State Color $v$ from $colList(v)$ chosen uniformly at random, say $c$
            \State Remove $v$ from its bucket and mark it as processed \label{line:remove1}%\MH{I think you want specific processed, $P[v]$} 
            \For{ $u \in adj(v)$} \LComment{Neighbors of $v$ in $G_c$}
                \If{ $u$ is uncolored and $c \in colList(u)$}
                    \State Remove $c$ from $colList(u)$ \label{line:remove-col}
                    \If{$colList(u)$ is empty}
                        \State Mark $u$ as processed
                        \State $V_u = V_u \cup u$
                        \State Continue
                    \EndIf
                    \State Remove $u$ from its current bucket \label{line:remove2}
                    \State Insert $u$ to $B[colList(u).size()]$
                \EndIf 
            \EndFor
        \EndWhile
    \end{algorithmic}
\end{algorithm}

Algorithm~\ref{alg:list-color-dyn-grd} stores the vertices of $G_c$ in an array of buckets $B$, according to the size of their color lists. The bucket at $B[i]$ holds the vertices $v \in V$, whose size of the $colList(v)$ is $i$. We define the \emph{lowest bucket} from the array as the non-empty bucket with the smallest index. The algorithm marks all the vertices as unprocessed and continue until all the vertices are processed. In each iteration, the algorithm finds the lowest bucket and chooses a vertex uniformly at random from it. This vertex is colored with an arbitrary color from its list and marked as processed. We say the color is $c$. Then, the algorithm scans all neighbors of this vertex in $G_c$ and removes $c$ if it exists in their color list. If any neighbor's list becomes empty, we mark this neighboring vertex as processed. 
%\MH{I think you want an array to mark processed for a specific vertex.}
The runtime of Algorithm~\ref{alg:list-color-dyn-grd} is $\mathcal{O}((|V_c|+|E_c|) \calL)$, since there might be at most $|V_c|$ iterations and in each iteration we require $\mathcal{O}(\calL)$ time to find the lowest bucket. The removals of a vertex from a bucket (Lines~\ref{line:remove1} and \ref{line:remove2}) can be implemented in constant time by an auxiliary array that stores the location of the vertex in the bucket, while the removal of a color from $colList$ (Line~\ref{line:remove-col}) takes $\mathcal{O}(\calL)$. Thus, processing a chosen vertex requires $\mathcal{O}(\delta(v) \calL)$ time, and summation over all vertices gives the total time.

\subsection{Analysis of the Algorithm}
Building on the analysis in~\cite{AssadiCK19}, we now show that, under reasonable assumptions on palette size, with a high probability our algorithm constructs in each iteration a conflict graph that is sublinear in the graph size. Following the previous section, we will omit the subscript $\ell$ since the results hold for any iteration. Recall that $G_c$ is the conflict subgraph computed from $G$ and $n \coloneqq |V|$. We will require the following concentration results.
\begin{lemma}[Chernoff-Hoeffding bound~\cite{Mitzenmacher-book-05}]
\label{lem:chernoff}
    Let $X_1,\ldots, X_m$ be $m$ independent binary random variables such that $Pr(X_i) = p_i$. Let $X = \sum_{i=1}^m X_i$  and $\mu = \bbE[X]$. Then the following holds for $0<\gamma\leq1$:
    \begin{align}
        Pr\left[X \geq (1+\gamma) \mu\right] \leq e^{\left(-\frac{\mu \gamma^2}{3}\right)}
    \end{align}
\end{lemma}

\begin{lemma}
\label{lemm:palcol}
At iteration, $\ell$, let $G(V,E)$ be the graph to be colored with Palette size $\calP$, and the color list size for each vertex $O(\log n)$. Let the average and maximum degree of $G$ be $\bard$ and $\Delta$, respectively. Then, the total number of colors found by Algorithm~\ref{alg:pal-col-meta} is $\sum_\ell \calP_\ell$. 
At iteration $\ell$ of the Algorithm~\ref{alg:pal-col-meta}, the following relations hold:
\begin{enumerate}[label={~\ref{lemm:palcol}.\arabic*)}]
    \item \label{lab:exp_deg} Expected degree of vertex $v \in G_c$ is $\mathcal{O}(\frac{\delta(v)}{\calP} \log^2{n})$.
    \item \label{lab:max_deg} Assuming $\frac{\Delta}{\calP} = \mathcal{O}(\log n)$, the maximum degree and the maximum number of edges in $G_c$ is $\mathcal{O}(\log^3{n})$ and $\mathcal{O}(n\log^3{n})$ respectively with high probability.
    \item Assuming $\frac{\bard}{\calP} = \mathcal{O}(\log n)$, the expected number of edges in $G_c$ is $\mathcal{O}(n \log^3 n)$.
\end{enumerate}
\end{lemma}
\begin{IEEEproof}
Due to the design of the algorithm, the total number of colors used is: $\sum_\ell \calP_\ell$. %For the following three proofs, we assume $n = |V|$.
%\begin{enumerate}[leftmargin=*, noitemsep, wide=1pt, topsep=2pt]
\begin{enumerate}[wide]
    \item Let us consider vertex $v$, and let $T$ be the size of the color list for $v$. Let $X_{v,u}$ be a binary random variable that takes $1$ if the edge $(v,u)$ is in the conflict graph $G_c$, and $0$ otherwise. Also, let $X_v = \sum_{u \in adj(v)} X_{v,u}$. Let us fix the colors in the list of $v$ as $c_1, c_2, \ldots c_T$. The probability that at least one of these colors is shared with a neighbor, $u$, i.e., $Pr(X_{v,u} = 1)$ is $\mathcal{O}(\frac{T}{\calP}$). So the expected degree of $v$ is $\bbE[X_v] = \sum_{u \in adj(v)} \mathcal{O}(\frac{T}{\calP}) = \mathcal{O}(\frac{\delta(v)}{\calP} \log^2{n})$, since $T = \mathcal{O}(\log^2 n)$.
    
    \item Since the maximum degree is $\Delta$, the maximum expected degree in $G$ is $\mathcal{O}(\frac{\Delta}{\calP} \log^2{n}) = \mathcal{O}(\log^3 n)$ according to Lemma~(\ref{lab:exp_deg} and our assumption on the ratio $\frac{\delta(v)}{\calP}$. We will now show the high probability concentration result. We note that all $X_{v,u}$s are independent of each other. So, using the Chernoff-Hoeffding bound (Lemma~\ref{lem:chernoff}) and setting $\gamma = 1$, the probability that $Pr[X_v \geq  2(\log^3 n)] \leq e^{-(\log^3 n)/3} \leq \mathcal{O}(n^{-\log^2 n})$. So with high probability the maximum degree of $v$ is $\mathcal{O}(\log^3 n)$ assuming the ratio $\frac{\Delta}{\calP} = \mathcal{O}(\log n)$.  
    \item We can further improve the bound by using the average degree ($\bard$) of $G$ rather than the maximum degree on our assumption. Let $Y$ be the random variable representing the sums of degrees of the graph at level $\ell$. From Lemma~(\ref{lab:exp_deg}, the expected sums of degrees,
    \begin{align*}
        \bbE[Y] =\sum_{v \in n} X_v = &\sum_{v \in V} \mathcal{O}\left(\frac{\delta(v)}{\calP} \log^2{n}\right)\\
        & = \sum_{v \in V} \mathcal{O}\left(\frac{\bard}{\calP} \log^2{n}\right) = \mathcal{O}(n \log^3 n).
    \end{align*}
    This relationship follows since the sums of degrees of a graph can be replaced with sums of average degrees. The final equality follows from our assumption, $\frac{\bard}{\calP} = \mathcal{O}(\log n)$.
    %$E[Y] = \sum_{v \in n} X_v = \sum_{v \in V} O(\frac{\delta(v)}{\calP_\ell} \log^2{n_\ell}) $
\end{enumerate}
\end{IEEEproof}

\begin{comment}

\subsection{Memory and Work Requirements}
The memory requirement of our implementation is $\mathcal{O}(|V|(\alpha\mathcal{L}+S+2)+2|E_c|)$. The worst-case work is $\mathcal{O}(\alpha\mathcal{L}(|V||V-1|+deg(G_c)|V_c|))$
\MH{Ferdous, please complete this part.}
\end{comment}
\section{Parallel GPU implementation}
\label{sec:gpu-graph-creation}
\begin{algorithm}
    \caption{GPU Conflict Graph Construction}
    \label{alg:gpu-imp}
    \begin{algorithmic}[1]
        \AlgIn{Pauli strings, $V$ and their list of colors $colList$}
        \AlgOut Conflict graph {$G_{c}$} in CSR format
        %\State $G \gets G_{orig}$
        \State AvailMem = min($2|V|(|V|-1)$, MaxAvailGPUMem)
        \State Allocate AvailMem on the GPU
        \State $V_{edgecount}, E_{coo}\gets$  build\_unordered\_coo(colList, $V$)
        \State $V_{offsets}\gets$ exclusive\_sum($V_{edgecount}$)
        \If{$|E_{coo}| \le $AvailMem/2}
            \State $G_c \gets$ generate\_csr\_gpu($V_{offsets}, E_{coo}$)
        \Else
            \State $G_c \gets$ generate\_csr\_cpu($V_{offsets}, E_{coo})$
        \EndIf
    \end{algorithmic}   
    \label{alg:palette-coloring}
\end{algorithm}
The conflict subgraph construction on Line~\ref{line:buildgraph} of Algorithm~\ref{alg:pal-col-meta} significantly dominates the execution time for our application. Because of this, we implemented a parallel GPU version to reduce the bottleneck. We note that, despite the original graph being dense, in almost all the cases the conflict graph is expected to be significantly sparse (details in \S\ref{subsec:quality-mem}). Although the pairwise comparisons of vertices are independent of each other, due to the unknown number of conflicting edges at runtime, we designed an implementation capable of generating the conflict graph in Compressed Sparse Row (CSR) format that is not only efficient in memory usage during construction but also enables contiguous access of memory for processing the conflict graph.  We present this implementation in \cref{alg:gpu-imp}. As a preprocessing (now shown in the \cref{alg:gpu-imp}), we  copy the input data on the GPU with a size of $N\mathcal{L}|V|/10$ 4-byte values for the encoded Pauli strings and their list of colors, and initialize $2|V|$ edge offset counters. We use 8 bytes for the counter if $|V|^{2}\ge2^{32}$; otherwise, the offset counters are 4-bytes. All remaining available memory on the GPU, or the worst-case edgelist size of $2|V|(|V|-1)$, whichever is smaller, is then allocated to store the unordered edgelist, and the conflict graph generation kernel is launched on the GPU. For this kernel, each thread processes one of the $|V|(|V|-1)/2$ possible edges and, and inspect whether the edge is both a complement and conflicting. If so, the edge is written in the output and the respective edge offsets are incremented. 
After the kernel execution, we are left with an unordered edge list of size $|E_c|$.%, assuming that each unordered edge is represented only once. 
The edges of the complement graph are determined independently during conflict graph construction, and the complement graph does not need to be stored on the GPU memory. Since for CSR representation each edge is stored twice, if $|E_c|$ used less than half of the available GPU memory, then we generate the CSR output on the GPU. Otherwise, we read the unordered edge list and convert it to CSR on the host CPU instead. We attempted warp-level reduction on the offset values to condense the number of global atomic operations, but the number of conflicting edges was much smaller than the total number of possible edges ($\leq$5\% in most cases), so the overhead outweighed the benefits.
% We also attempted to directly address the upper triangular matrix, reducing empty loop checks by implementing an integer square root to calculate the proper row and column from a given edge index.
To get an estimation for $|E_c|$ to preallocate memory on the GPU, we developed a machine learning based predictor, which we describe next (\S\ref{sec:Prediction}). 
%The predictor is designed to estimate the parameters that provide the best trade-offs to meet the quality and memory/runtime goals.

% \textit{\textbf{Some quick notes, needs to be fleshed out:}} Discuss parallel filtering of complement + conflicting edges, conversion to unordered edgelist, then CSR post-processing. Uses encoding scheme introduced in \cref{sec:pauliencoding}. Grid-stride loops for 100\% theoretical GPU occupancy. Atomic reduction attempted, but sparsity of conflicting edges ($\le{5\%}$ of all edge comparisons) resulted in more overhead than benefits. Uses direct index addressing to operate on the upper block triangle matrix (sans diagonal). Main bottleneck is uncoalesced global accesses for comparing colors, as each thread processes its own list at a stride of $L$ or more, where $L$ is the list size.
\section{Prediction of Palette Size}
\label{sec:Prediction}

% \subsection{Methodology}
We employ a machine learning (ML) based methodology to predict the palette size  $\mathcal{P}$ and $ \alpha$ values to simultaneously minimize the number of final colors $\mathcal{C}$ and the number of conflicting edges $|E_c|$. As these two objectives are conflicting, we introduce $\beta$ as the weighting factor to determine the balance between minimizing $\mathcal{C}$ and $|E_c|$. The goal for this prediction is to find the optimal combination of $(\mathcal{P}, \alpha )$ to minimize $(\beta \cdot \mathcal{C} + (1 - \beta) \cdot |E_c| )$, i.e.,
\begin{align}
    \min_ {( \mathcal{P}, \alpha )} (\beta \cdot \mathcal{C} + (1 - \beta) \cdot |E_c| ).
\label{eq:ml}
\end{align}
We generate a dataset by varying values for percentile palette size, $\mathcal{P}'=\frac{\mathcal{P}}{|V|} \times 100$ (as a percentage of the number of vertices $|V|$ in the complement graph $G^\prime$) and $ \alpha$ to perform a grid search. We capture the $(\mathcal{P}', \alpha )$ combinations that minimizes Eq. (\ref{eq:ml}) for different values of $\beta$. Finally, this dataset trains regression models to predict the $( \mathcal{P}', \alpha )$ for a given graph $G$ and $\beta$. We train the regressor with several molecules and test its accuracy on a new set of molecules. The methodology can be summarized as follows:

%\begin{itemize}
\begin{itemize}[leftmargin=*, noitemsep, topsep=2pt]
    \item \textit{Step 1:} For a given graph $G(V,E)$, perform sweeps on $( \mathcal{P}', \alpha )$ and compute $(\mathcal{C}, E_c)$. 

    \item \textit{Step 2:} For a particular value of $\beta$, compute objective in Eq. \eqref{eq:ml}, and select the optimal choice of $( \mathcal{P}', \alpha )_{opt}$.

    \item \textit{Step 3:} Run Step 2 for different values of $\beta$, and collect corresponding values of $( \mathcal{P}', \alpha )_{opt}$ to construct the data-set for the graph $G$.

    \item \textit{Step 4:} Run Steps 1-3 for the graphs correspond to different molecules to construct the complete training set. 

    \item \textit{Step 5:} Train the regressor model with $( \mathcal{P}', \alpha )$ as outputs of the model for a given graph $G$ and $\beta$. 

    \item \textit{Step 6:} After the model is trained, we provide a new graph and a particular choice of $\beta$, for which, the model predicts an optimal choice of $( \mathcal{P}', \alpha )$ that optimizes for Eq. \eqref{eq:ml}.
    
\end{itemize}
\noindent \textbf{Model Training and Results:}
We generated a dataset for the molecules provided in \Cref{tab:dataset} for percentile palette sizes $\mathcal{P}' \in \{1\%,2.5\%,5\%....,20\%\}$ and $\alpha \in \{0.5,1.0,....,4.5\}$. We capture the $\langle \mathcal{P}', \alpha \rangle$ combinations that minimized Eq. (\ref{eq:ml}) for $\beta \in \{0.1,0.2,....,0.9\}$. %We trained the models by randomizing the input graph and keeping $80\%$ of the data for training and $20\%$ for testing.
From the dataset, we used the first five molecules for training and the last two for testing the regression analysis.
We experimented with several linear (ridge, lasso) and nonlinear predictors (svm-kernel-rbf, decision trees, random forests)~\cite{Russell2009}.
%and observed that the nonlinear regression models performed better for the prediction task. 
For given input of: ($\beta, \ V, \ E)$, the nonlinear regression models performed better in predicting the ideal $\langle \mathcal{P}', \alpha \rangle$ combination. 
In particular, the {\em random forest regressor} provided the best performance with multiple iterations, with a mean absolute percentage error (MAPE) of 0.19 and an R-squared value of 0.88 over 100 iteration runs. We selected the number of trees (estimators) to be 100 and the maximum tree depth of 20. 
This methodology provides \pic{} a means to predict optimal choice of palette and list sizes for an input with values for ($\beta, \ V, \ E)$, where $\beta$ provides the trade-off. 
We note that while our model is trained specifically for the dense input graphs used in this work, it can be extended to any family of inputs that can be well characterized with data as described above.

\ignore{
\subsection{Pareto Optimal Points}

We require Pareto optimal points (where no change can improve one objective without deteriorating the other to achieve the optimal trade-off solutions for the two conflicting objectives. Fig.~\ref{fig:pareto} shows the Pareto optimal points for the midsize molecules from the dataset as given in Table~\ref{tab:dataset }. The normalized final color, $\mathcal{C}$, ranges between 0.125-0.3 for the normalized conflicting edges, $E_c$, that ranges between 0.01-0.85 in Fig.~\ref{fig:pareto}.%\sayak{statement not clear, rewrite needed}.
The knee of this Pareto curve stays between 0.135 to 0.225 for normalized $\mathcal{C}$ and 0.05 to 0.4 for normalized $E_c$. The corresponding normalized $\mathcal{P}$ ranges between 0.14-0.21, and $\alpha$ ranges between 1-6 for the knee of the Pareto curve. The Pareto curve gets smoother with a more granular set of values for $\mathcal{P}$ and $\alpha$. The user can leverage $\beta$ for assigning different weights for the objectives suitable for his intended application and subsequently extract the optimal values of $\mathcal{P}$ and $\alpha$. %\sayak{can you please change the terms 'norm' in fig caption to 'normalized'?}

%We experimented with several algorithms and found that random forest method provides the best predictions. We train the model by randomizing the input and keeping $80\%$ of the data for training and $20\%$ for validation. We test the model with new inputs that were not used during training. The model predictions for a sample graph is illustrated in Fig.\ref{}, for different values of $\beta$. \textsl{Note: indicate the parameters used in the training (e.g., number of trees, tree depth, etc.)}

%%%%%%%%%%%%%%%%%%%%%%%%%%%%%%%%%%%%
\begin{figure}[!htb]
\centering
\includegraphics[width=0.5\textwidth]{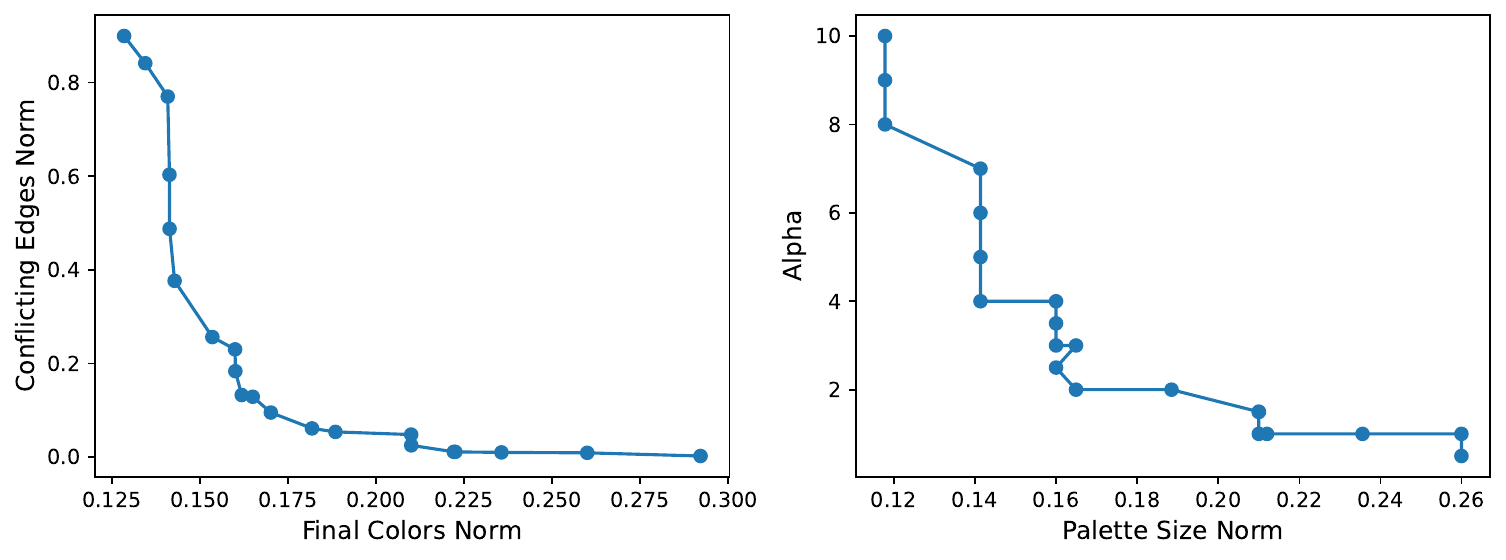} 
\caption{(left) Pareto optimal points for the two objectives, (right) and the corresponding Normalied $\mathcal{P}$ and $\alpha$.}
\label{fig:pareto} 
\end{figure}
%%%%%%%%%%%%%%%%%%%%%%%%%%%%%%%%%%%%\
}
\section{Experimental Evaluation}
\label{sec:ExptResults}

% \subsection{Dataset and Experimental Setup}
% \subsection{Dataset and Experimental setup}
%Please add the following packages if necessary:
%\usepackage{booktabs, multirow} % for borders and merged ranges
%\usepackage{soul}% for underlines
%\usepackage[table]{xcolor} % for cell colors
%\usepackage{changepage,threeparttable} % for wide tables
%If the table is too wide, replace \begin{table}[!htp]...\end{table} with
%\begin{adjustwidth}{-2.5 cm}{-2.5 cm}\centering\begin{threeparttable}[!htb]...\end{threeparttable}\end{adjustwidth}
\begin{table}[!b]\centering
\caption{Molecule Dataset in our experiment}\label{tab:dataset}
\scriptsize
\begin{tabular}{l|rrr}\toprule
Molecule Name & \# of qubits &\# of Pauli terms &\# of edges \\\midrule
%H2\_sto3g &17 &72 \\
%H2\_631g &89 &1988 \\
%H2\_6311g &225 &12688 \\
%H4\_3D\_sto3g &449 &51,168 \\
%H4\_2D\_sto3g &657 &109,224 \\
%H4\_1D\_sto3g &689 &120,264 \\
%\midrule
H6\_3D\_sto3g & 12 &8,721 &19,178,632 \\
H6\_2D\_sto3g & 12 &18,137 &82,641,188 \\
H6\_1D\_sto3g & 12 &19025 &90,853,544 \\
H4\_2D\_631g & 16 &22529 &127,024,320 \\
H4\_3D\_631g & 16 &34481 &297,303,496 \\
H4\_1D\_631g & 16 &42449 &450,624,984 \\
H4\_2D\_6311g & 24 &154641 &5,979,614,600 \\
\midrule
H4\_3D\_6311g & 24 &245089 &15,017,722,736 \\
H8\_2D\_sto3g & 16 &271,489 &18,513,622,112 \\
H8\_1D\_sto3g & 16 &274,625 &18,944,162,720 \\
H4\_1D\_6311g & 24 &312817 &24,464,823,272 \\
H8\_3D\_sto3g & 16 &419,457 &44,149,092,736 \\
H6\_3D\_631g & 24 &554,713 &77,027,619,060\\
H10\_3D\_sto3g & 20 &1,274,073 &410,446,230,804 \\
\midrule
H6\_2D\_631g & 24 &2,027,273 &1,028,164,570,684 \\
H6\_1D\_631g & 24 &2,066,489 &1,068,358,440,628 \\
H10\_2D\_sto3g & 20 &2,093,345 &1,108,417,973,696 \\
H10\_1D\_sto3g & 20 &2,101,361 &1,116,895,244,280 \\
\bottomrule
\end{tabular}
\end{table}
We evaluate the performance and quality of the solution obtained using \pic{} to the current state-of-the-art approaches. For our evaluation, we used a machine equipped with a 64-core AMD EPYC 7742 CPU, 512GB host DDR memory, and an NVIDIA A100 GPU with 40GB of HBM memory.
%\mm{@Bo: please fill the blanks (...).}
\Cref{tab:dataset} lists the datasets we selected for our evaluation. % These datasets represent a wide range of hydrogen H$_n$ ($n=2,4,6,8,10$) 1D/2D/3D molecular systems in increasing size and dimension, and they have been selected because ... . 
These systems are chosen with a deliberate intent to encompass a broad spectrum of quantum scenarios. By picking $H_n$ molecular systems with varying values of $n = 4, 6, 8, 10$, we aim to ensure diversity in system size, ranging from simple to complex structures, enabling us to gauge the algorithm's scalability and performance across different magnitudes. The incorporation of three spatial configurations for each $H_n$ molecular system, namely 1D, 2D, and 3D,  introduces dimensional variability, shedding light on the tool's capability to manage problems with different geometric complexities and symmetries. As the size of the $H_n$ system increases, so does the intricacy of electron-electron interactions and correlations. By integrating systems of different sizes, the goal is to critically assess the tool's proficiency in addressing varying degrees of electron correlation, a pivotal aspect in quantum calculations. Moreover, this diverse selection, spanning across multiple system sizes and dimensions, serves as an effective stress test for \pic{}. It not only offers insights into its performance benchmarks but also aids in pinpointing potential areas of enhancement, ensuring a comprehensive evaluation of its reliability and robustness.
We classify our datasets into three categories:
\begin{enumerate*}[label=\textit{\roman*)}]
    \item Small ($\leq 10$ Billion edges);
    \item Medium ($\leq 1$ Trillion edges); and
    \item Large ($> 1$ Trillion edges).
\end{enumerate*}
We evaluate our implementation by considering:
\begin{itemize}%[leftmargin=*, noitemsep, topsep=2pt]
%\begin{itemize}
    \item \textbf{Performance:} We present a detailed performance study of the execution time of \pic{} and compare it against two of the state-of-the-art GPU implementations of the distance-1 coloring: 
    \begin{enumerate*}[label=\textit{\roman*)}]
       \item Kokkos-EB: edge-based coloring included as part of the kokkos-kernels~\cite{deveci2016parallel,Bogle2020,bogle2022parallel}, and  
       \item ECL-GC-R: shortcutting and reduction based heuristics for JP-LDF~\cite{alabandi2020increasing,Alabandi2022}. 
    \end{enumerate*} 
    \item \textbf{Quality:} We assess the quality of coloring obtained by \pic{} with respect to sequential greedy coloring implementations in ColPack~\cite{Gebremedhin2005,Gebremedhin2013}, and GPU implementations of Kokkos-EB, and ECL-GC-R. We also study the quality versus memory trade-offs in these implementations.
    \item \textbf{Parameter Sensitivity:} We study the impact of palette size ($\calP$) and color list size ($\alpha, \calL$) on the final coloring, the number of conflicting edges, and the runtime of \pic{}. 
    %We also study the effectiveness of a random forest predictor for predicting the initial values of the parameters. 
\end{itemize}

We limit our relative comparisons only to the small dataset due to limitations imposed by specific implementations. In fact, ColPack and Kokkos-EB run out of memory beyond the small datasets (Kokkos-EB also runs out of memory for the last instance of the small dataset). ECL-GC-R does not support a graph size larger than what a 32-bit integer type can represent (2 to 4B). All results presented here are averaged over five runs. Five different seeds for pseudo-random number generation are used for \pic{} runs. %Maximum resident set size during the execution of the program is used for reporting the memory required.
For \pic{}, we present the results using only the greedy list coloring of \cref{alg:list-color-dyn-grd} to color the conflict graph since it provided better coloring relative to the static ordering algorithms. 

We note that \pic{} does not require loading the entire graph into memory to color it.  Instead, it computes the conflicting subgraph on-the-fly at every iteration, thus providing the memory improvement.  This is fundamentally different from the previous state-of-the-art algorithms. In fact, ColPack, Kokkos-EB, and ECL-GC-R require loading of the entire graph into memory before coloring it. Therefore, we decided to be conservative in comparing the performance of the different implementations, and we \emph{include} the conflict subgraph construction time for \pic{} while we \emph{exclude} the graph construction time for all the other approaches. Under these conservative settings, we show that \pic{} is better or comparable to ECL-GC-R and within a factor $2\times$ slower than Kokkos-EB. 
We also note that the explicit construction of a complement graph is expensive for large instances.

The two main parameters of \pic{} are the size of palettes ($\calP$) and the color list ($\calL$). At any iteration of the algorithm with $V$ as the vertices considered, in our experiments, $\calP$ represents the percentage of vertices, and $\calL = \alpha \log{|V|}$, where $\alpha$ is the coefficient. We omit the subscript since, in each iteration, the same percentage value and $\alpha$ are used. We report the number of colors, running time in seconds, and maximum resident set size in GB for memory. Apart from these, we also define a few other metrics as follows.
\begin{itemize}%[leftmargin=*, noitemsep, topsep=2pt]
    \item \textbf{Color percentage:} percentage ratio between the number of colors to the number of vertices of the input, i.e., $\frac{\calC}{|V|}*100$. It represents the percentage of shrinkage of the Pauli strings to the unitaries (impact on the application). 
    \item \textbf{Maximum Conflicting Edge percentage:} The percentage ratio between the maximum number of conflicting edges (across all iterations of \pic{}) to the number of complement edges of the graph, i.e., $\frac{|E_c|}{|E|}*100$.
    % This percentage represents the compaction achieved by \pic{} compared to state-of-the-art applications which require storing all complement edges.
\end{itemize}
%For all the algorithms compared, we report running time in seconds and maximum resident set size in GB for memory.  

%We note that unlike the existing coloring packages, which requires the graph to be loaded into memory before coloring it, \pic{} do not load the whole graph. Instread at every iteration, it computes the conflicting subgraph on the fly, thus providing the memory improvement. For reporting the runtime,\pic{} includes this graph build times. But for other algorithms, we \emph{do not} include the graph construction time from the molecular dataset. Yet, we show that \pic{} for many choices of parameters are better or comparable than ECL-GC-R and within $2\times$ slower than Kokkos-EB.
% We describe our dataset in Table~\ref{tab:dataset }. We test all runs on a system equipped with a 64-core AMD EPYC 7742, 512GB host RAM, and a 40GB NVIDIA A100 GPU. Each experiment is repeated with five different seeds, and three runs per seed. Each seed leads to the same coloring result, regardless of CPU- or GPU-side processing, so multiple runs per seed only slightly varies in execution time.

\subsection{Quality and Memory Comparisons}
\subsubsection{Small Dataset}
\label{subsec:quality-mem}

\begin{table}[!htp]
    \caption{Quality comparisons of the algorithms. Results in bold are the best coloring. Norm.: $\calP = 12.5\%, \alpha = 2$; Aggr.: $\calP=3\%, \alpha=30$.}\label{tab:quality-compare}
    \scriptsize
    \resizebox{\columnwidth}{!}{
    \begin{tabular}{l|rrrr|rr|rr}\toprule
    Problem &\multicolumn{4}{c|}{ColPack} &\multicolumn{2}{c|}{\pic{}} &Kokkos-EB &ECL-GC\\\cmidrule{1-9}
    &LF &SL &DLF &ID & Norm.& Aggr.& & \\\midrule
    H6\_3D\_sto3g &2479 &902 &901 &952 &1425.4 &\dag\textbf{880.6} &1040.17 &943 \\
    H6\_2D\_sto3g &5389 &1598 &\textbf{1580} &1634 &2901.5 &\dag 1587.4 &1749.17 &1596 \\
    H6\_1D\_sto3g &5771 &1672 &\textbf{1601} &1689 &3036.1 &\dag 1650.4 &1815.83 &1642 \\
    H4\_2D\_631g &10049 &1922 &\textbf{1694} &1917 &3579.8 &1784.2 &1772.67 &1860 \\
    H4\_3D\_631g &15883 &2729 &2633 &2668 &5431.4 &\dag\textbf{2606} &2478.50 &2596 \\
    H4\_1D\_631g &19412 &3241 &\textbf{2943} &3233 &6538 &3212.8 &3426.17 &3098 \\
    H4\_2D\_6311g &72493 &8615 &\textbf{6944} &8628 &22463.8 &8917.4 &NA &NA \\
    \bottomrule
    \end{tabular}
    }
\end{table}
\begin{comment}
\begin{table}[htp]
    \caption{Memory comparison of the algorithms: Maximum resident memory in GB}\label{tab:mem-compare}
    \scriptsize
    \resizebox{\columnwidth}{!}{
    \begin{tabular}{l|r|rr|rr}\toprule
    Problem & ColPack &\multicolumn{2}{c|}{Picasso} & & \\\cmidrule{1-2}\cmidrule{3-6}
    & & Norm. & Aggr. &Kokkos-EB &ECL-GC-R \\\midrule
    H6\_3D\_sto3g &0.38 &\textbf{0.08} &0.23 &1.19 &0.30 \\
    H6\_2D\_sto3g &1.52 &\textbf{0.16} &0.90 &4.50 &0.77 \\
    H6\_1D\_sto3g &1.68 &\textbf{0.17} &0.97 &4.93 &0.83 \\
    H4\_2D\_631g &2.72 &\textbf{0.20} &1.24 &6.81 &1.10 \\
    H4\_3D\_631g &5.66 &\textbf{0.31} &3.10 &15.69 &2.37 \\
    H4\_1D\_631g &10.77 &\textbf{0.38} &4.31 &23.69 &3.51 \\
    H4\_2D\_6311g &140.23 &\textbf{2.06} &57.12 &NA &NA \\
    \bottomrule
    \end{tabular}
    }
    \end{table}
\end{comment}
\Cref{tab:quality-compare,tab:mem-compare} show the number of colors achieved and the memory requirements of the compared algorithms for the small dataset, respectively. We experimented with four ordering heuristics for sequential greedy coloring that are commonly considered in the literature: Largest First Degree (LF), Smallest Degree Last (SL), Dynamic Largest Degree First (DLF), and Incidence Degree (ID). The previous study on unitary partitions~\cite{izmaylov2019unitary,peng2022mapping} only considered LF greedy coloring algorithm. We refer you to the excellent survey of \citet{Gebremedhin2005} for details on vertex orderings.

\cref{tab:quality-compare} reports the average number of colors over five runs for all the algorithms. We show results from two different configurations of \pic{}:
\begin{enumerate*}[label=\textit{\roman*)}]
    \item Normal: $\calP = 12.5\%, \alpha = 2$ and
    \item Aggressive: $\calP = 3\%, \alpha = 30$.
\end{enumerate*}
We observe that \pic{} always produces fewer colors than the LF heuristic using \pic{}'s normal mode. We find that the DLF heuristics provide the best coloring. However, the aggressive configuration of \pic{} provides coloring that is within {\bf 5\%} in $4/7$-th of the inputs (marked with $\dag$ in \cref{tab:quality-compare}), and within {\bf 10\%} in all cases except the largest of the small dataset.  Finally, we observe that ECL-GC-R provides better coloring than Kokkos-EB on $4/6$-th of the inputs for which they could compute a solution. In both cases, the coloring provided by \pic{}'s aggressive configuration is better or within $5\%$ of what is obtained through ECL-GC-R and Kokkos-EB.

% For memory, we show the memory requirement of the heuristic that takes the smallest memory. Please see the survey~\cite{Gebremedhin2005} for a detailed description of the greedy ordering. For \pic{}, we show the average number of colors and corresponding average memory on two configurations: i) normal mode ( $\calP = 12.5\%, \alpha = 2$) and ii) aggressive mode ($\calP = 3\%, \alpha = 30$). We also show the quality achieved and memory required by Kokkos-EB and ECL-GC-R. The previous study on unitary partitions~\cite{izmaylov2019unitary,peng2022mapping} only considered LF greedy coloring algorithms. We can see that \pic{} can always achieve better coloring than the LF heuristics using very relaxed parameters (normal mode). In general, the DLF heuristics provide the best coloring. But with aggressive parameter choice, \pic{} can achieve very close to the sequential DLF coloring results. ECL-GC-R provides better color than Kokkos-EB. 
\begin{table}[htp]
    \caption{Memory comparison of the algorithms: Maximum resident memory in GB}\label{tab:mem-compare}
    \resizebox{\columnwidth}{!}{
    \begin{tabular}{l|r|rr|rr}\toprule
    Problem & ColPack &\multicolumn{2}{c|}{Picasso} & & \\\cmidrule{1-2}\cmidrule{3-6}
    & & Norm. & Aggr. &Kokkos-EB &ECL-GC-R \\\midrule
    H6\_3D\_sto3g &0.38 &\textbf{0.08} &0.23 &1.19 &0.30 \\
    H6\_2D\_sto3g &1.52 &\textbf{0.16} &0.90 &4.50 &0.77 \\
    H6\_1D\_sto3g &1.68 &\textbf{0.17} &0.97 &4.93 &0.83 \\
    H4\_2D\_631g &2.72 &\textbf{0.20} &1.24 &6.81 &1.10 \\
    H4\_3D\_631g &5.66 &\textbf{0.31} &3.10 &15.69 &2.37 \\
    H4\_1D\_631g &10.77 &\textbf{0.38} &4.31 &23.69 &3.51 \\
    H4\_2D\_6311g &140.23 &\textbf{2.06} &57.12 &NA &NA \\
    \bottomrule
    \end{tabular}
    }
\end{table}

\Cref{tab:mem-compare} shows the maximum resident memory in GB of the reference implementations for the small dataset.
We find that the normal configuration of \pic{} is the most memory efficient. In particular, it requires {\bf 68$\times$} lesser memory than the ColPack for H4\_2D\_6311g while Kokkos-EB and ECL-GC-R run out of memory and couldn't compute a solution for the same instance. We observe between $14\times$ and $60\times$ lower memory utilization when comparing \pic{}'s normal mode to Kokkos-EB, and 
a reduction in memory usage of $\approx5\times$ when considering \pic{}'s aggressive mode. ECL-GC-R shows to be more memory efficient than Kokkos-EB and is comparable to \pic{}'s aggressive mode. However, ECL-GC-R memory optimizations come with runtime penalties that we will detail in our performance evaluation (\cref{ssec:PerfEval}).

\subsubsection{Medium and Large Dataset}
\label{sec:med-large-quality}
%%%%%%%%%%%%%%%%%%%%%%%%%%%%%%%%%%%%
\begin{figure}[!htb]
\centering
\input{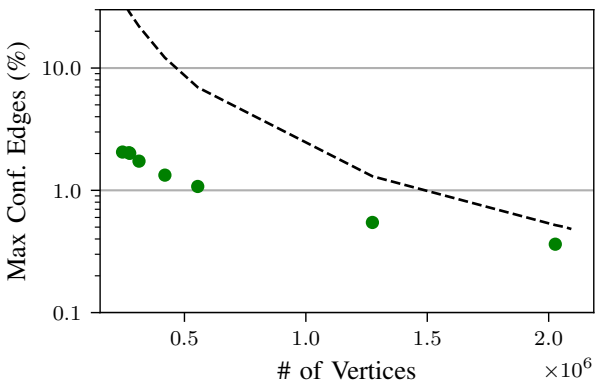} 
\caption{Input dataset scaling on the iterative GPU implementation up to 2 million vertices. \textit{$\alpha=2$, $\mathcal{P}=12.5\%$}. The black dashed line in the top plot denotes the maximum conflicting edge ratio supported by a 40GB NVIDIA A100.}
\label{fig:mem-medium} 
\end{figure}
As mentioned earlier, none of the compared algorithms could generate coloring for these inputs; we will only discuss results obtained from \pic{}. Using $\calP=12.5\%$ and $\alpha=2$, we could color all the medium inputs. We even observe a slight quality improvement compared to the small dataset. For the chosen parameters, the color percentage is {\bf 14--15\%} (as a percent of $|V|$), whereas for the same parameter setting, \pic{} achieved 14--16\% for the smaller dataset (see \Cref{tab:quality-compare}). For the four large inputs (with over 1 Trillion edges), we set $\calP=12.5\%$ as before, but changed $\alpha = 1$. We were then able to generate coloring for all the large instances with this parameter except the largest one, which ran out of GPU memory. For the first three inputs of the large dataset, \pic{} achieved a color percentage of {\bf 16.2--16.4\%}. These results suggest that \pic{} can achieve reasonably high-quality coloring even for larger datasets, for which none of the current state-of-the-art GPU implementations included in our study were able to color within the 40 GB of available GPU memory of our system. Due to the quadratic scaling of the complement edges relative to the number of Pauli strings, an increasingly smaller conflicting-edge ratio is needed to satisfy memory requirements for larger problems. This can be observed in \cref{fig:mem-medium} where the black dashed line traces the limits of the maximum fraction of conflict edges ($\%$) that can fit the A100 for each input. We address the memory issue by choosing more conservative parameters ($\calP$ and $\alpha$) as demonstrated for the largest inputs in our dataset.

\subsection{Performance Evaluation}
\label{ssec:PerfEval}
\subsubsection{CPU-only Vs. GPU-assisted Implementation} We report the speedup of our GPU implementation over the CPU-only implementation of \pic{} in \Cref{tab:cpu-gpu-comp}. We show the average time over five runs for the conflict graph construction and the total time of the CPU-only implementation in the $2^{nd}$ and $3^{rd}$ columns, respectively. Here, the reported conflict graph construction time includes the cumulative time spent in building the conflict graphs during each iteration of the algorithm. The process of building conflict graph accounts for over $98\%$ of execution time (geo. mean) for these problems. We accelerate the conflict graph build on GPUs (\Cref{sec:gpu-graph-creation}). The last two columns of \Cref{tab:cpu-gpu-comp} report the speedup of our GPU implementation with respect to the conflict graph build step and the total runtime. We see that as the problem size increases, the speedup also increases, and we expect that trend to continue for even larger problems. We report results only for the small datasets because we used a cut-off time of 1 hour, and the CPU implementation was able to complete only the small dataset within that time budget. The geometric mean of the speed up for the conflict graph construction step is $\sim${\bf 60$\times$}, and that results into a $\sim${\bf 16$\times$} speed up for the entire application (geo. mean). We note that our GPU implementation produces exactly the same coloring as the CPU-only one because the conflict graph construction is deterministic.
\begin{comment}
\begin{table}[!htp]\centering
\caption{Conflict graph creation time comparison between sequential and gpu implementation of \pic.}\label{tab:time_comp}
\resizebox{\columnwidth}{!}{
\begin{tabular}{lrrrr}\toprule
Problem &CPU Time (s) &GPU Time (s) &Speedup \\\midrule
H6\_3D\_sto3g &3.14 &0.13 &24.37 \\
H6\_2D\_sto3g &14.87 &0.34 &43.76 \\
H6\_1D\_sto3g &16.35 &0.36 &45.36 \\
H4\_2D\_631g &24.47 &0.47 &51.55 \\
H4\_3D\_631g &57.67 &0.79 &73.21 \\
H4\_1D\_631g &91.02 &1.09 &83.85 \\
H4\_2D\_6311g &1,428.94 &8.24 &173.36 \\
\bottomrule
\end{tabular}
}
\end{table}
\end{comment}

\begin{table}[!htp]\centering
\caption{Runtime comparison for CPU only and GPU assisted implementation. $\calP = 12.5\%, \alpha=2$}\label{tab:cpu-gpu-comp}
\scriptsize
\begin{tabular}{l|rr|rr}\toprule
&\multicolumn{2}{c|}{CPU only} &\multicolumn{2}{c}{GPU assisted} \\\cmidrule{2-5}
Problem &Graph Build &Total &Graph Build &Total \\%\cmidrule{1-5}
&Time(s) &Time(s) &Speedup &Speedup \\\midrule
H6\_3D\_sto3g &3.14 &3.26 &24.37 &2.21 \\
H6\_2D\_sto3g &14.87 &15.19 &43.76 &8.94 \\
H6\_1D\_sto3g &16.35 &16.69 &45.36 &9.75 \\
H4\_2D\_631g &24.47 &24.92 &51.55 &13.41 \\
H4\_3D\_631g &57.67 &58.40 &73.21 &26.30 \\
H4\_1D\_631g &91.02 &92.00 &83.85 &35.25 \\
H4\_2D\_6311g &1,428.94 &1,436.10 &173.36 &110.90 \\
\midrule
Geo. Mean & & &59.54 &15.98 \\
\bottomrule
\end{tabular}
\end{table}

\subsubsection{Performance on Medium and Large Dataset}
%%%%%%%%%%%%%%%%%%%%%%%%%%%%%%%%%%%%
\begin{figure}[!htb]
\centering
\input{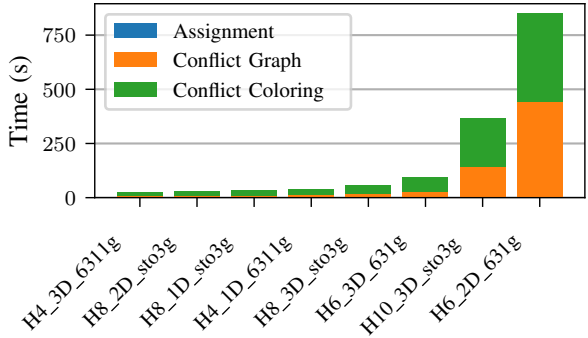} 
\caption{Input dataset scaling on the iterative GPU implementation up to 2 million vertices. \textit{$\alpha=2$, $\mathcal{P}=12.5\%$}.}
\label{fig:runtime-med-large} 
\end{figure}
%%%%%%%%%%%%%%%%%%%%%%%%%%%%%%%%%%%%
\Cref{fig:runtime-med-large} shows the running time (with a breakdown in components) for all the medium and one of the large datasets for our GPU implementation. The problems are sorted from left to right, with the smallest problem on the left. For all the inputs, we set $\calP = 12.5\%$, and $\alpha = 2$. We see that the conflict coloring that happens in CPU dominates the runtime. Despite this, we were able to color the largest graph with over 1 Trillion edges within {\bf 800 seconds} with a color percentage ranging between 14--15\%. For detailed quality results, see  \Cref{sec:med-large-quality}.

\begin{comment}
\subsection{Memory Assessment}
\input{IPDPS24/Floats/Table/mem-compare}
\Cref{tab:mem-compare} shows the memory requirements for each application. Note that \pic{} can have standard (left, palette size ratio = 12.5\%, $\alpha=2$) and aggressive (right, palette size ratio = 3\%, $\alpha=30$) configurations, with tradeoffs between quality and memory requirement/runtime in between. As can be seen, \pic{}'s standard configuration has the smallest memory requirement, while its most aggressive is similar to ECL-GC-R. Colpack and Kokkos-EB both have much larger memory requirements compared to \pic{} and ECL-GC-R.
% Comparison of memory footprint. We can present some results on $\beta$ parameter sweeps. \textit{Do we want to show this in a plot or table?}
\end{comment}
\subsection{Performance Comparison with Kokkos-EB and ECL-GC-R}
%%%%%%%%%%%%%%%%%%%%%%%%%%%%%%%%%%%%
\begin{figure*}[!htb]
\centering
\input{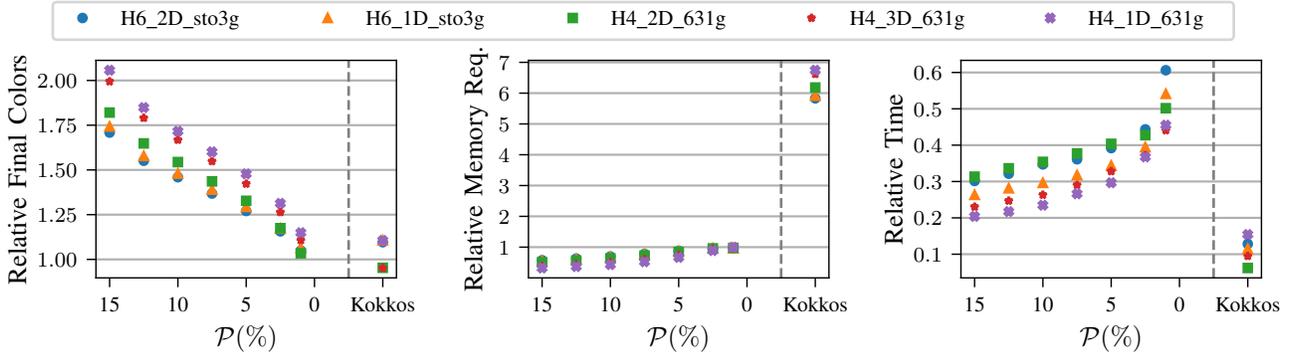} 
\vspace{-2mm}
\caption{Performance comparison of \pic{} and Kokkos-EB on the small datasets relative to ECL-GC-R execution time. \textit{For \pic{} runs, $\calP$ is varied and $\alpha=4.5$}}
\label{fig:palette-sweep-kokkos} 
\vspace{-5mm}
\end{figure*}

%%%%%%%%%%%%%%%%%%%%%%%%%%%%%%%%%%%%
\Cref{fig:palette-sweep-kokkos} compares \pic{} to the current state-of-the-art GPU implementations of graph coloring: Kokkos-EB and ECL-GC-R. The study is limited to the small dataset due to memory constraints imposed by specific implementations. We study quality of solution in terms of number of colors produced by the implementations, memory usage, and execution time.  The results in \Cref{fig:palette-sweep-kokkos} are normalized with respect to ECL-GC-R.  We fixed $\alpha=4.5$ and we varied $\calP$ from $1\%$ up to $15\%$ in our \pic{} runs. We observe that the quality of solution achieved by \pic{} increases when reducing the palette size ($\calP$). When $\calP=1\%$, \pic{} matched the quality of solution of Kokkos-EB and ECL-GC-R (within $5\%\text{--}15\%$).

\cref{fig:palette-sweep-kokkos} shows that ECL-GC-R produces the best quality results at the expense of a much longer execution time. In comparison, \pic{}'s runs with $\calP=1\%$ completed between $0.44\times$ and $0.60\times$ the time used by ECL-GC-R. Kokkos-EB approach results in the fastest execution time: between $0.06\times$ and $0.15\times$ the time of ECL-GC-R. However, Kokkos-EB uses between $5.83\times$ and $6.74\times$ the memory used by ECL-GC-R while \pic{} had comparable or reduced memory usage ($0.99\times \text{-- } 0.32\times$) with respect to the same baseline.

\subsection{Parameter Sensitivity}
%%%%%%%%%%%%%%%%%%%%%%%%%%%%%%%%%%%%
\begin{figure}[!htb]
\centering
\input{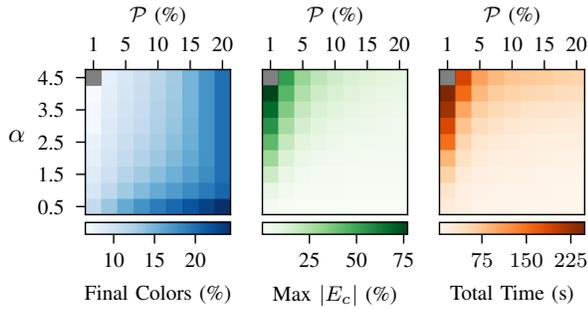}
\caption{Impact of $\mathcal{P}$ and $\alpha$ using H4\_2D\_6311g on final colors, number of conflicting edges and runtime for different inputs (lighter color is better).}
\label{fig:ParameterSweep} 
\vspace{-0.7mm}
\end{figure}
%%%%%%%%%%%%%%%%%%%%%%%%%%%%%%%%%%%%
The heatmap in \cref{fig:ParameterSweep} shows the impact of $\calP$ and $\alpha$ on the final colors, conflicting edges, and execution time for a representative input (H4\_2D\_6311g).  The heatmaps show normalized quantities with respect to what is observed for the same input graph. Therefore, as a measure of quality we report the fraction of colors with respect to the number of vertices in the input (lower is better) while we report execution time and the normalized fraction of the total number of conflicting edges processed by the algorithm as measures of the work done by the algorithm (lower is better).

The general trend favors smaller $\calP$ and larger $\alpha$ to achieve a lower number of final colors at the cost of extra work. Conversely, larger $\calP$ and smaller $\alpha$ lead to lower conflicting edges, and therefore, lower memory requirements and lead to faster execution time. We observed similar trends on all our datasets. These observations led us to design the approach in \Cref{sec:Prediction} to jointly minimize work and the size of the coloring by controlling $\calP$ and $\alpha$.

% \subsection{Machine Learning Model Results (Accuracy)}
% We generated a dataset for the molecules provided in \Cref{tab:quality-compare} for percentile palette size $\mathcal{P}' \in \{1\%,2.5\%,5\%....,20\%\}$ and $\alpha \in \{0.5,1.0,....,4.5\}$. We capture the $\langle \mathcal{P}', \alpha \rangle$ combinations that minimized Eq. (\ref{eq:ml}) for $\beta \in \{0.1,0.2,....,0.9\}$. We used the first five molecules for training and the last two for testing the regression models. Given the small number of predictors ($\beta, \ V, \ E)$, the nonlinear regression models performed better in predicting the ideal $\langle \mathcal{P}', \alpha \rangle$ combination. Specifically, the random forest regressor performed best with a mean absolute percentage error (MAPE) of 0.19 and an R-squared value of 0.88 over 100 iteration runs. We selected the number of trees (estimators) to be 100 and the maximum tree depth.
\section{Conclusions and Future Work}
\label{sec:Conclusions}
%\MH{SM, please feel free to trim/cut the first paragraph. What Bo wrote in the second paragraph is awesome!}

%The wave function and the system Hamiltonian play crucial roles in quantum simulations aimed at elucidating physical and chemical phenomena, where the wave function is responsible for state preparation, and the system Hamiltonian controls the evolution of the state.
%The processes of state preparation and evolution are conducted through unitary operations, which are rewritten as a combination of Pauli strings each being a unitary. However, such techniques suffer from the so-called curse of dimensionality in large-scale applications. 
%A near-term approach to address the curse of dimensionality and the associated quantum measurement overhead is to seek a compact unitary representation for both the Hamiltonians and the wave function generators through {\em clique partitioning} of the graph representing Pauli strings.  The clique partitioning of a graph is equivalent to the coloring of it's complement graph. We presented \texttt{Picasso}, a scalable graph coloring algorithm that enables the computation of large-scale clique partitions in dense (about 50\% density) graphs, a first-of-its-kind memory-efficient parallel algorithm targeting accelerator platforms. 
We demonstrated the memory efficiency of \texttt{Picasso} using a large set of inputs and compared its performance with state-of-the-art approaches for graph coloring. In the realm of quantum computing, the introduction of \pic{} marks a notable advancement. It is, quite plausibly, the inaugural scalable graph algorithm and implementation tailored for Hamiltonian and wave function partitioning that surpasses the capabilities of contemporary state-of-the-art quantum emulators. What amplifies its significance is its versatile nature; the same tool can be adeptly employed in qubit tapering, thereby reducing the effective number of qubits required for a given problem. When synergized with other methodologies, such as single reference guidance~\cite{peng2022mapping}, \pic{} promises the ability to solve systems comprising 100 to 1000 spin orbitals (i.e. qubits). It is within these vast and complex systems that the much-vaunted quantum advantage is believed to manifest, positioning our tool at the forefront of a transformative computational frontier.

Our future work will focus on developing distributed multi-GPU parallel implementations along with new algorithms for predicting $\mathcal{P}$ and $\alpha$ values to enable better trade-offs for quality and performance. We plan to further optimize our algorithm to address inputs from diverse applications with varying degrees of sparsity. We also plan to develop efficient algorithms for clique partitioning that explore applications beyond quantum computing. To the best of our knowledge, this is the first of its kind study for computing coloring of dense graphs on limited memory accelerator platforms and believe that it will enable several applications that critically depend on the computation of clique partitioning and graph coloring, as well as enable the development of memory-efficient graph algorithms.

\section*{Acknowledgement}
The research is supported by the Laboratory Funded Research and Development at the Pacific Northwest National Laboratory (PNNL), the U.S. DOE Exascale Computing Project's (ECP) (17-SC-20-SC) ExaGraph codesign center at PNNL, and NSF awards  to North Carolina State University. 

 %\bibliographystyle{IEEEtran}
 %\bibliography{IPDPS24/refs,IPDPS24/hala}
\printbibliography
\end{document}